# Reliability-considered Multi-platoon's Groupcasting using the Resource Sharing Method

Chung-Ming Huang[1], Yen-Hung Wu [1], and Duy-Tuan Dao[2]

[1]Dept. of Computer Science and Information Engineering, National Cheng Kung University, Tainan, Taiwan
{huangcm, wuyh}@locust.csie.ncku.edu.tw
[2]Faculty of Electronics and Telecommunications Engineering, The University of Danang-University of Science and Technology, Vietnam
ddtuan@dut.udn.vn

*Abstract*— **In the context of 5G platoon communications, the Platoon Leader Vehicle (PLV) employs groupcasting to transmit control messages to Platoon Member Vehicles (PMVs). Due to the restricted transmission power for groupcasting, it may need to pick one PMV as the Platoon Relay Vehicle (PRV) to be responsible for re-groupcasting messages of PLVs. To optimize the usage of limited spectrum resources, resource sharing can adopted to enhance spectrum efficiency within the platoon. This study proposes a resource allocation method, which is called Resource Sharing for Platoon Groupcasting (RSPG), for platoon's groupcasting based on transmission reliability. RSPG utilizes the tripartite matching to assign a subchannel to either a PLV or PRV that shares the assigned subchannel with the corresponding individual entity (IE), which does not belong to any platoon. The simulation results shown that the proposed method has the better performance in terms of the QoS satisfaction rate of IEs, the number of allocated subchannels for platoons, and spectral efficiency.**

*Keywords*: **Vehicle Platoon, Multi-platoon Communications, Platoon Groupcasting, Tripartite Matching, Resource Allocation, Resource Sharing.**

## I. INTRODUCTION

Platooning involves a group of vehicles coordinating their speeds and distances while moving together [1]. A platoon consists of a Platoon Leader Vehicle (PLV), which is essentially positioned at the front of the platoon, and several Platoon Member Vehicles (PMVs), which follow the driving instructions of the PLV. Platooning can lead to benefits such as reduced fuel consumption, alleviated traffic congestion, improved traffic flow efficiency, and enhanced road safety [2]. 5G cellular network's platoon groupcast [3], which is like multicast, is a way to send information to a specific group of vehicles. It aims to provide efficient and reliable communication between the PLV and PMVs to ensure timely exchange of critical information for a platoon [3]. In the context of platooning, groupcast allows the PLV to groupcast control information to all of the PMVs in the platoon simultaneously.

The signal transmitted by the PLV may suffer from path loss fading when the platoon becomes too long, which results in poor reception for PMVs that are far away the PLV. Platoon Relay Vehicle (PRV) refers to the use of a certain PMV as the relay to extend the communication range and enhance signal reception within the platoon. A PRV can receive PL's groupcasted messages and re-groupcast them to the PMVs, which are outside PLV's groupcasting range, to ensure PLV's messages to be received by the longer distance away PMVs. The relaying mechanism enhances the overall reliability and coverage of platooning. Many papers considered the relay issue and proposed their methods based on different objectives [4~7]. Several resource management and allocation methods that address various technical concerns have been proposed for the communication in a platoon. For example, (1) the subchannels allocated to PLVs for broadcasting messages should be orthogonal to the subchannels assigned to the PMVs [8]; (2) the subchannels allocated to the PLV for groupcasting messages can be reused by uplinked subchannels of individual entities (IEs), which denote smartphone users and free vehicles that don't join any platoon [8][9].

This work proposes a resource management and allocation method called Resource Sharing for Platoon Groupcasting (RSPG) that considers reliability of PLVs' and PRVs' groupcasting on the condition of sharing resource with IEs because platoons always coexist with IEs using limited spectrum resource. The proposed RSPG method is devised to maximize intra-platoon reliability while meeting the Quality of Service (QoS) requirements of IEs. Using the proposed RSPG method, a tripartite matching problem is formulated to allocate subchannels that maximize platoon reliability while considering the QoS constraints of IEs by optimizing the transmitted power of PLV/PRVs.

The remaining part of the paper is organized as follows: Section 2 presents related works. Section 3 presents the system model for the considered 5G NR C-V2X platoon-based network. Section 4 mathematically formulates the reliability problems for both PLV/PRVs and PMVs. Section 5 presents the proposed algorithm that is for resolving the formulated problems. Section 6 presents the performance analysis and the comparison with other methods. Finally, Section 7 has conclusion remarks and highlights potential future work.

## II. RELATED WORK

In [4], the authors focused on improving the dissemination of cooperative awareness messages (CAMs) from the PLV in the platooning scenario using a combined approach of relay selection and power control. The goal is to select an optimal relay vehicle to enhance the channel conditions while ensuring reliable communication links through power control. Two methods proposed in [4] for resolving the formulated problem are (i) the centralized method, which resolves the optimization of relay selection and power control by considering large scale signal variations, and (ii) the distributed method, which combines theoretical analysis from the centralized method with the predecessor-leader following (PLF) control strategy in a sequential communication mode. The simulation results shown that the proposed method effectively ensure link reliability even in the situation with the low transmission power.

In [5], the authors study the V2X network design for platooning, which focuses on the significant impact of RSU-based packet relaying on the inter-vehicle distance. Two types of resources for relaying are as follows: (i) Relying on licensed spectrum, which uses the C-V2X technology. This type has the advantage of using a dedicated spectrum in a scheduled mode such that the relay link is collision exempt because transmissions are scheduled. (ii) Relying on unlicensed spectrum, for which the Road Side Unit (RSU) is considered to be able to overhear the packets sent by the PLV and then retransmit these packets in a broadcasting manner using the IEEE 802.11p system. According to the simulation results, the proposed methods reduce failures and inter-vehicle distances with the minimal delay. Simulation results demonstrate the lowest packet loss rate using the licensed spectrum, but a cost-performance compromise favors RSU-based relay using the unlicensed spectrum. However, the RSU-based packet relay costs more resource than using the vehicle relay and it needs more power to send messages to RSU.

In [6], the authors developed a Markov model to analyze different types of communication links (intra-vehicle and vehicle-to-RSU). Then, the authors proposed a control and communication system designed for platooning. The proposed method evaluates the performance using two relaying strategies: (1) C-V2V relaying and (2) RSU relaying. According to the simulation results, the proposed C-V2V relaying scheme significantly reduces the distance between vehicles compared to two other methods, i.e., one has no relaying and the other one uses RSU relaying.

In [7], the authors proposed a collaborative platoon communication and control design that consists of two communication phases: (i) PLVs' information dissemination, for which the corresponding subchannels only can be used by the PLVs' and PRVs' broadcasting; (ii) PMVs' information dissemination, for which PMVs use the same timeslots to exchange information with neighboring PMVs. The proposed method utilizes PRVs to forward PLV's messages using a broadcasting approach in the first phase to extend the communication range of the PLV. In the second phase, which is for PMVs' information dissemination, if the number of available subchannels is smaller than the number of wireless links required for PMVs' communication, the proposed method allows intra-platoon PMVs' resource sharing. That is, one subchannel can be allocated to multiple intra-platoon PMVs to improve spectrum utilization. The selection of relay vehicles is formulated as an integer programming problem, and a dynamic programming method was proposed to solve the selection problem. For PMVs' resource allocation, an adaptive distributed model predictive control was proposed. According to the simulation results, the reliability of broadcasting PLV's information is improved, and the scale is increased through the proposed relay selecting method. However, the proposed method only focuses on the scenario of a single platoon and does not consider the scenario of multiple platoons.

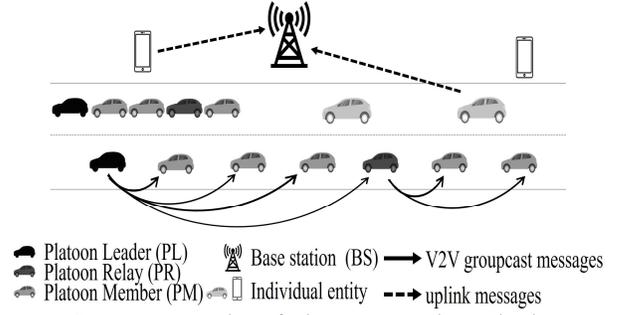

**Figure 1.** An example of the proposed method's system configuration.

## III. The System Model

This Section describes details of the system model, including network model, communication model, and channel model.

The network model includes (i) one Base Station (BS), (ii) $M$ platoons, in which a platoon can have one PLV, some PMVs and one optional PRV, and (iii) $C$ IEs. An example of the proposed system configuration is depicted in Figure 1.

Let there be $V_m$ platoon vehicles (PVs) in platoon $m$, where vehicles are numbered from 0 to $V_m - 1$ starting from the PLV, i.e., $PV_0$ is the PLV and $PV_1$ to $PV_{V_m-1}$ are PMVs, which can also be denoted as $PM_1$ to $PM_{V_m-1}$, respectively. Let $R$ be PLV's groupcasting range, which covers PMVs 1 to $n_m$, i.e., PMVs 1 to $n_m$ can directly receive the groupcasted messages sent from PLV. For the PMVs that are outside PLV's groupcasting range, $PM_{n_m}$ is selected as the PRV to re-groupcast its received messages sent from PLV to PMVs $PM_{n_m+1}$ to $PM_{V_m-1}$. If $n_m = V_m - 1$, it indicates that the PLV of platoon $m$ can groupcast its messages to all of its PMVs; however, if $n_m < V_m - 1$, the PLV of platoon $m$ cannot groupcast its messages to all PMVs and thus vehicle $PM_{n_m}$, which is the farthest PMV in the range of PLV's groupcasting, can play the PRV role to re-groupcast PLV's messages to $PM_i$, $i = n_m + 1,.., V_m - 1$. The adopted communication model is as follows. Let the network bandwidth be divided into K orthogonal subchannels denoted by the set $\mathbb{K} = \{1, 2, \ldots, K\}$ and each subchannel $k$ contain several resource blocks (RBs). The subchannel's usage principle adopted in this work is as follows: (i) a PLV or a PRV cannot share its subchannel with any PMV but can share its subchannel with one IE's uplinked subchannel. (iii) No subchannel sharing among IEs. (iv) An IE can share its subchannel with a PLV/PRV.

The adopted channel model is as follows. The channel power gain $h_{i,j}^m$ from transmitting vehicle $i$ to receiving vehicle $j$ in platoon $m$ is calculated as follows:

$$h_{i,j}^m = G * \left(d_{i,j}^m\right)^{-\alpha} * (h_0^m)^2 = \beta_{i,j}^m * \bar{h}_{i,j}^m \qquad (1)$$

where G denotes the power gain constant introduced by communication equipment, $h_0^m \sim CN(0,1)$ denotes a complex gaussian random variable representing Rayleigh fading, $d_{i,j}^m$ denotes the distance from platoon m's vehicle $i$ to platoon m's vehicle $j$, $\alpha$ denotes the path loss exponent, $\beta_{i,j}^m$ is a random variable describing the channel gain's uncertainty, $\bar{h}_{i,j}^m$ is the statistical average channel gain. The matrix $Y^k = [y_{m,g}^k]_{\forall m,g} \in \{0,1\}^{M*2}$, in which 2 denotes PLV and PRV, represents the assignment of subchannel $k$ to PLV's or PRV's groupcasting of platoon $m$, where $y_{m,g}^k = 1$ indicates the vehicle of platoon $m$ uses subchannel $k$ for groupcasting, $g = 0$ denotes PLV's groupcasting and $g = 1$ denotes PRV's groupcasting; otherwise, $y_{m,g}^k = 0$. The matrix $Z = [z_c^k]_{\forall c} \in \{0,1\}^{C*K}$, in which $K$ is the number of subchannels and $C$ is the number of IEs, $z_c^k = 1$ represents the assignment of subchannel $k$ to IE $c$. The SINR that PLV ($i = 0$) or PRV ($i = 1$) $i$ groupcasts messages to PMV $j$ over subchannel $k$ is represented as follows:

$$SINR_{i,j,k}^m = \frac{P_i^m * h_{i,j}^m}{\sigma^2 + I_{i,k}^m} \qquad (2)$$

where $P_i^m$ denotes the transmitted power of platoon $m$'s PLV/PRV, $h_{i,j}^m$ denotes the channel power gain from the transmitting PLV/PRV to the receiving PMV $j$ in platoon $m$, $\sigma^2$ denotes the power of the additive white Gaussian noise (AWGN), $I_{i,k}^m$ denotes the interference from the IE $c$, which shares subchannel $k$ with the PLV/PRV of platoon $m$. $I_{i,k}^m$ is represented as follows:

$$I_{i,k}^m = \sum_{c=1}^{C} z_c^k * P_c^k * h_{c,i,k}^m \qquad (3)$$

subject to

$$\sum_{c=1}^{C} z_c^k = 1 \qquad (3a)$$

where $z_c^k$ represents the assignment of subchannel $k$ to IE $c$, $P_c^k$ denotes the transmitted power of IE $c$, $h_{c,i,k}^m$ represents the channel gain from IE $c$ to the PLV/PRV of platoon $m$ over subchannel $k$. Equation $(3a)$ denotes that a PLV's/PRV's groupcasting subchannel can only be shared with one IE. The SINR from transmitting IE $c$ to BS using subchannel $k$ is as follows:

$$SINR_{c,k} = \frac{P_c * h_c}{\sigma^2 + I_{c,k}^m} \qquad (4)$$

where $P_c$ denotes the transmitted power of IE $c$, $h_c$ represents the channel gain from IE $c$ to BS, $I_{c,k}$ denotes the total co-channel interference from a PLV's groupcasting, a PRV's groupcasting, or some PMVs' unicasting that share subchannel $k$ with IE $c$. $I_{c,k}$ is as follows:

$$I_{c,k} = \sum_{m=1}^{M} \sum_{g=0}^{1} y_{m,g}^k * P_g^k * h_{g,c,k}^{mg} \qquad (5)$$

subject to

$$0 \leq P_g^k \leq P_{max} \qquad (5a)$$

$$\sum_{m=1}^{M} \sum_{g=0}^{1} y_{m,g}^k \leq 1 \qquad (5c)$$

where $y_{m,g}^k$ represents the assignment of subchannel $k$ to PLV's or PRV's groupcasting link of platoon $m$, $P_g^k$ denotes the transmitted power of PLV's/PRV's groupcasting over subchannel $k$, $h_{g,c,k}^{mg}$ represents the (groupcasting) channel gain from platoon m's PLV ($g = 0$) or PRV ($g = 1$) to IE $c$ over subchannel $k$, Equation (5a) denotes the transmitted power should not exceed the maximum threshold, Equation (5c) means that an IE can share its allocated subchannel with at most one PLV's or one PRV's groupcasting. The receiving bit rate of the vehicle $j$ of platoon $m$ from the transmitting PLV/PRV $i$ of platoon $m$ over subchannel $k$ can be expressed as follows:

$$R_{i,j,k}^m = log_2(1 + SINR_{i,j,k}^m) \qquad (6)$$

The receiving bit rate in the BS from the transmitting IE $c$ over subchannel $k$ can be expressed as follows:

$$R_{c,k} = log_2(1 + SINR_{c,k}) \qquad (7)$$

## IV. PROBLEM FORMULATION

In this Section, the problem of maximizing reliability for PLV's/PRV's groupcasting is formulated.

Achieving the reliability of a communication link can be done by guaranteeing the successful transmission's probability from vehicle $i$ to vehicle $j$ of platoon $m$, i.e., the corresponding link's SINR is greater than the pre-defined SINR threshold, which can be expressed as follows:

$$Pr_{i,j}^m = Prob(SINR_{i,j,k}^m \geq \gamma_{thr}) \geq \theta_{th} \qquad (8)$$

where $\gamma_{thr}$ is the pre-defined SINR threshold and $\theta_{th}$ is the pre-defined probability threshold. Let $Pr_{i,j}^m$ be the successful transmission probability from vehicle $i$ to vehicle $j$ in platoon $m$. Then, the transmission reliability $Rel$-$g_j^m$ from PLV/PRV to PMV $j$ in platoon $m$ of the single-relay-vehicle's situation[1] is as follows:

$$Rel\text{-}g_j^m =$$
$$\begin{cases} Pr_{0,j}^m, \text{when PM vehicle } j \text{ is in PLV's groupcasting range.} \\ Pr_{r,j}^m, \text{when PM vehicle } j \text{ is only in PRV's groupcasting range.} \end{cases} \quad (9)$$

The objective is to find the PRV in each platoon that can maximize the reliability. For each platoon, the PRV selection problem can be modeled as follows:

$$\max_{r^m} \sum_{j=1}^{\nu_m-1} Rel\text{-}g_j^m \qquad (10)$$

where $r^m$ is the selected relay vehicle for platoon $m$. Based on Equations (8) and (9), Equation (10) can be transformed to the following one:

$$\max_{r^m} \sum_{j=1}^{\nu_m-1} Prob(SINR_{i,j,k}^m \geq \gamma_{thr}) \;, \; i\epsilon\{0,r^m\} \qquad (11)$$

Then, referring Equation (2), Equation (11) can be transformed to the following one:

$$\max_{r^m} \sum_{j=1}^{\nu_m-1} \left( Prob\left(\frac{P_i^m * h_{i,j}^m}{\sigma^2 + I_{i,k}^m} \geq \gamma_{thr}\right) \right) \qquad (12)$$

Then, Equation (12) can be further transformed to the following one:

---

1. In this work, the maximum transmitted power of PL and PR vehicle's groupcasting is set to be able to cover at least half of platoon vehicles in a platoon with the maximum interference caused by individual entity because of resource sharing, i.e., the proposed method's groupcasting can always cover all platoon vehicles in a platoon with at most one PR vehicle.

$$\max_{r^m} \sum_{j=1}^{\mathcal{V}_m-1} \left( Prob \left( \frac{P_i^m * \beta_{i,j}^m * \bar{h}_{i,j}^m}{\sigma^2 + I_{i,k}^m} \geq \gamma_{thr} \right) \right) \quad (13)$$

subject to

$$1 \leq n_m \leq V_m - 1 \qquad (13a)$$

$$1 \leq r^m \leq n_m \qquad (13b)$$

$$0 \leq P_i^m \leq P_{max} \qquad (13c)$$

$$\sum_{m=1}^{M} \sum_{g=0}^{1} y_{m,g}^k \leq 1 \qquad (13d)$$

$$\sum_{c=1}^{C} z_c^k \leq 1 \qquad (13e)$$

$$SINR_{c,k} \geq \delta_{thr} \qquad (13f)$$

$$SINR_{i,j,k}^m \geq \gamma_{thr} \qquad (13g)$$

where $\bar{h}$ denotes the statistical average channel gain, $\beta$ denotes the random variable describing the channel gain's uncertainty. In Equation (13), constraint (13a) means that the last PMV $n_m$ inside PLV's groupcasting range of platoon $m$ should be one of platoon m's composed vehicle, (13b) means that platoon $m$'s PRV is in the groupcasting range of platoon $m$'s PLV, (13c) denotes that (i) the transmitted power of groupcasting should not exceed the maximum threshold, (13d) means that a subchannel $k$ can be used by at most one groupcasting vehicle, (13e) means that a subchannel $k$ can be used by at most one IE, (13f) is the QoS requirement of each IE, and (13g) is the QoS requirement of each groupcasting. The successful transmission probability in Equation (13) can be re-written as follows:

$$\max_{r^m} \sum_{j=1}^{\mathcal{V}_m-1} \left( Prob \left( \beta_{i,j}^m \geq \frac{\gamma_{thr} * (\sigma^2 + I_{i,k}^m)}{P_i^m * \bar{h}_{i,j}^m} \right) \right) \quad (14)$$

Since $\beta_{i,j}^m$, which describes the channel gain's uncertainty, has exponential distribution with a mean of 1, which follows the probability density function $f(x)$:

$$f(x) = \begin{cases} e^{-x}, \text{for } x \geq 0 \\ 0 \ , \text{for } x < 0 \end{cases} \quad (15)$$

According to the Fundamental theorem of Calculus:

$$\int_a^b e^{-x} \, dx = (-e^{-b}) - (-e^{-a}) \qquad (16)$$

Based on Equation (16), Equation (14) can be rewritten as follows:

$$Prob \left( \beta_{i,j}^m \geq \frac{\gamma_{thr} * (\sigma^2 + I_{i,k}^m)}{P_i^m * \bar{h}_{i,j}^m} \right) = \int_{\frac{\gamma_{thr} * (\sigma^2 + I_{i,k}^m)}{P_i^m * \bar{h}_{i,j}^m}}^{\infty} e^{-x} \, dx$$

$$= (-e^{-\infty}) - \left( -e^{-\frac{\gamma_{thr} * (\sigma^2 + I_{i,k}^m)}{P_i^m * \bar{h}_{i,j}^m}} \right)$$

$$= 0 + e^{-\frac{\gamma_{thr} * (\sigma^2 + I_{i,k}^m)}{P_i^m * \bar{h}_{i,j}^m}}$$

$$= e^{-\frac{\gamma_{thr} * (\sigma^2 + I_{i,k}^m)}{P_i^m * \bar{h}_{i,j}^m}} \qquad (17)$$

Therefore, referring to Equations (14) and (17), the objective function depicted in Equation (13) can be transformed to the following Equation:

$$\max_{r^m} \sum_{j=1}^{\mathcal{V}_m-1} e^{-\frac{\gamma_{thr} * (\sigma^2 + I_{i,k}^m)}{P_i^m * \bar{h}_{i,j}^m}} \ , \ i \epsilon \{0, r^m\} \qquad (18)$$

subject to

$$0 \leq P_i^m \leq P_{max} \qquad (18a)$$

Since (1) the SINR requirement $\gamma_{thr}$, (2) the power of the AWGN $\sigma^2$, and (3) the statistical average channel gain $\bar{h}$ in Equation (18) are fixed, Equation (18) is equal to the following one:

$$\min_{r^m} \left( \sum_{j=1}^{\mathcal{V}_m-1} \frac{I_{i,k}^m}{P_i^m} \right), \ i \epsilon \{0, r^m\} \qquad (19)$$

## V. The Proposed Method

In this Section, the proposed method for solving the formulated problems introduced in Section IV is presented.

Equation (19) is solved as follows: (i) one subchannel will be assigned to a platoon for PLV's groupcasting and (ii) the other subchannel will be assigned to a platoon for PRV's groupcasting optionally, which depends on the existence of the PRV in a platoon:

$$\min_{r^m} \left( \sum_{j=1}^{\mathcal{V}_m-1} \frac{I_{i,k}^m}{P_i^m} \right), \ i \epsilon \{0, r^m\}$$

$$= \min_{r^m} \sum_{j=1}^{\mathcal{V}_m-1} \frac{P_{c_i}^k * h_{c_i,i,k}^m}{P_i^m}, \ i \epsilon \{0, r^m\} \qquad (20)$$

where $c_i$ is the IE that shares subchannel $k$ with PLV/PRV $i$.

At most two subchannels, i.e., one for PLV's groupcasting and one for PRV's groupcasting, which may be shared with IEs, will be allocated to a platoon to yield the minimum result of Equation (20). The problem depicted in Equation (20) is equal to solve the tripartite matching problem, i.e., matching (i) an allocated subchannel, (ii) a PLV or a PRV and (iii) an IE together. Let groupcasting vehicle $g$ share its subchannel with IE $c$, the QoS requirement for IE $c$ can be formulated as follows:

$$SINR_{c,k} = \frac{P_c * h_c}{\sigma^2 + I_{c,k}} \geq \delta_{thr} \Rightarrow I_{c,k} \leq \frac{P_c * h_c}{\delta_{thr}} - \sigma^2 \quad (21)$$

According to Equations (4) and (21), the maximum transmitted power of the groupcasting vehicle can be calculated as follows:

$$P_g^{m_g} * h_{g,c,k}^{m_g} \leq \frac{P_c * h_c}{\delta_{thr}} - \sigma^2$$

$$\Rightarrow P_g^{m_g} \leq \frac{\frac{P_c * h_c}{\delta_{thr}} - \sigma^2}{h_{g,c,k}^{m_g}} \qquad (22)$$

The objective of the proposed RSPG method is to get the matching result of the tripartite matching problem, for which the devised algorithm is depicted in Algorithm 1.

Algorithm 1, which is called "the RSPG algorithm" hereafter, is explained as follows. Line 2 calls function *PL-IE-CHMatching* to find all candidate matchings, which are put in $\mathbb{T}$, for PLV's groupcasting. Function *PL-IE-CHMatching* derives the corresponding PLV's transmitted power when the IE's SINR constraint, i.e., constraint (h) in Equation (13), is

---

**Algorithm 1:** Resource Sharing for Platoon Groupcasting (RSPG)

● Input: $\mathbb{C}$ individual entities, $\mathbb{K}$ subchannels and $\mathbb{M}$ platoons.

● Output: resource allocation matching set $\mathbb{S}$.

1. $\mathbb{T} \leftarrow \{\}$;
   // $\mathbb{T}$ temporarily stores available matching $\{(c, k, m, g), x\}$, where (1) $(c, k, m, g)$ denotes the matching of having individual entity $c$ and platoon PL ($g$ =0) or PR ($g$ =1) vehicle of platoon $m$ to share subchannel $k$ and (2) $x$ denotes the upper bound of the transmitted power.
2. $\mathbb{T} \leftarrow PL\text{-}IE\text{-}CHMatching(\mathbb{C}, \mathbb{K}, \mathbb{M})$;
3. $\overline{\mathbb{T}} \leftarrow Sort(\mathbb{T}, x)$;
   //Sort elements in $\mathbb{T}$ descendingly based on $x$.
4. $\mathbb{S} \leftarrow \{\}$;
5. $\mathbb{S} \leftarrow ResultedMatching(\overline{\mathbb{T}}, 0)$;
6. $\mathbb{R} \leftarrow [R_1, R_2, \dots, R_M] = [-1, -1, \dots, -1]$
   //$\mathbb{R}$ is used to store the PR vehicle's index of each platoon.
7. **for** $m = 1 : M$ **do**:
8.    $p \leftarrow \{x' | \{(c', k', m, 0), x'\} \in \mathbb{S}\}$;
9.    $i \leftarrow 1$;
      //$i$ is used to temporary store the index of the farthest PM vehicle that is in PL's groupcasting range.
10.   $f \leftarrow false$;
      //$f$ is used to check whether the farthest PM vehicle in PL's groupcasting range is found or not.
11.   **While** $i \leq V_m - 1$ and $f = false$ **do**:
12.      calculate the SINR of the $i$-th PM vehicle using Equation (2).
13.      **if** $SINR_{0,i,k'}^m > \gamma_{th}$ **then**:
14.         $i \leftarrow i + 1$;
15.      **else**:
16.         $\mathbb{R}[m] \leftarrow i - 1$;
17.         $f \leftarrow ture$;
18.      **end if**;
19.   **end while**;
20. **end for**;
21. $\mathbb{T} \leftarrow \{\}$;
22. $\mathbb{T} \leftarrow PR\text{-}IE\text{-}CHMatching(\mathbb{C}, \mathbb{K}, \mathbb{M})$;
23. $\overline{\mathbb{T}} \leftarrow Sort(\mathbb{T}, x)$;
   //Sort elements in $\mathbb{T}$ descendingly based on $x$.
24. $\mathbb{S} \leftarrow \mathbb{S} + ResultedMatching(\overline{\mathbb{T}}, 1)$;
25. **return** $\mathbb{S}$;

---

**Function** $PL\text{-}IE\text{-}CHMatching$ ($\mathbb{C}, \mathbb{K}, \mathbb{M}$)

1. $\mathbb{T} \leftarrow \{\}$
2. **foreach** $k$ in $\mathbb{K}$ **do**:
      **foreach** $c$ in $\mathbb{C}$ **do**:
3.    **for** $m = 1 : M$ **do**:
4.       compute the SINR of individual entity $c$ by using Equation (4).
5.       **if** constraints (h) of Equation (13) is satisfied **then**:
6.          $x \leftarrow \min(P_{max}, \frac{\frac{P_c \cdot h_c}{\delta_{thr}} - \sigma^2}{h_{0,c,k}^m})$;
            //calculate the upper bound of PL's transmitted power.
7.          $\mathbb{T} \leftarrow \mathbb{T} + \{(c, k, m, 0), x\}$;
8.       **end if**;
9.    **end for**;
10.  **end foreach**;
11. **end foreach**;
12. **return** $\mathbb{T}$;

---

**Function** $ResultedMatching$ ($\overline{\mathbb{T}}, g$)

1. $\mathbb{S} \leftarrow \{\}$;
2. $i_{\overline{\mathbb{T}}} \leftarrow 0$;
3. //$i_{\overline{\mathbb{T}}}$ is used to store the index of $\overline{\mathbb{T}}$'s elements.
4. **while** $(i_{\overline{\mathbb{T}}} < |\overline{\mathbb{T}}|)$ **or** $(\mathbb{M} \neq \emptyset)$ **do**:
5.    **if** $c \in \mathbb{C}$ and $k \in \mathbb{K}$ and $m \in \mathbb{M}$ in $\overline{\mathbb{T}}[i_{\overline{\mathbb{T}}}]$ **then**:
6.       $\mathbb{S} \leftarrow \mathbb{S} + \overline{\mathbb{T}}[i_{\overline{\mathbb{T}}}]$;
         //allocate subchannel $k$ to entity $c$ and PL (PR), where input parameter $g = 0$ (1), vehicle of platoon $m$.
7.       $\mathbb{C} \leftarrow \mathbb{C} \setminus c$;
8.       $\mathbb{K} \leftarrow \mathbb{K} \setminus k$;
9.       $\mathbb{M} \leftarrow \mathbb{M} \setminus m$;
10.   **end if**;
11.   $i_{\overline{\mathbb{T}}} \leftarrow i_{\overline{\mathbb{T}}} + 1$;
12. **end for**;
13. **return** $\mathbb{S}$;

---

The initial value of $R_i$ is set as $-1$, which is initialized in Line 6 of the RSPG algorithm and means that PLV's groupcasting range of platoon $m$ can cover all of platoon $m$'s PMVs and thus the corresponding platoon $m$ does not need to find PRV. Lines 7 to 20 of the RSPG algorithm find PRV for each platoon. Line 8 of the RSPG algorithm finds the transmitted power of platoon $m$'s PLV. Lines 11 to 19 of the RSPG algorithm find the farthest PMV in PLV's groupcasting range, for which the perceived PL's SINR of the corresponding PMV is still greater than the threshold $\gamma_{th}$. Lines 13 to 18 of the RSPG algorithm check whether the $i^{th}$ PMV's SINR constraint is satisfied or not; if the answer is negative, then (i) the PMV whose index is $i-1$ is the PRV of platoon $m$ and (ii) the checking for the remaining PMVs is stop; otherwise, it continues to check the next PMV. Lines 21 to 25 of the RSPG algorithm add all available candidate matchings that can satisfy IE's SINR constraint to set $\mathbb{T}$. Line 22 of the RSPG algorithm calls function $PR\text{-}IE\text{-}CHMatching$ to find all candidate matchings for PRV's groupcasting.

Function $PR\text{-}IE\text{-}CHMatching$ derives the corresponding PRV's transmitted power when IE's SINR constraint, i.e., constraint (h) in Equation (13), is satisfied, for which the smaller value of (i) the PLV's derived transmitted power using Equation (22) and (ii) the maximum transmitted power $P_{max}$ is assigned (on Line 6) because the transmitted power cannot be satisfied, for which the smaller value of (i) the PLV's derived transmitted power using Equation (22) and (ii) the maximum transmitted power $P_{max}$ is assigned because the transmitted power cannot be higher than the maximum power $P_{max}$. Line 3 sorts the elements in set $\mathbb{T}$, i.e., all candidate matchings for PLV's groupcasting, based on PLV's transmitted power, i.e., the value of $x$, from high to low, i.e., descendently. Line 5 calls function $ResultedMatching$ to find all resulted matchings for PLV's groupcasting.

Function $ResultedMatching$ iteratively picks up the candidate matching that has the $i_{\overline{\mathbb{T}}}^{th}$ highest transmitted power: If entity $c$, PLV (PRV) of platoon $m$ and subchannel $k$ have not been matched, then add the matching to set $\mathbb{S}$; then remove $c$, $k$ and $m$ from $\mathbb{C}$, $\mathbb{K}$ and $\mathbb{M}$ respectively in Lines 7 to 9 of function $ResultedMatching$ because they have been allocated. Line 6 of the RSPG algorithm initiates vector $\mathbb{R}$, in which $R_i, i = 1..M$, stores the index of the PRV.

---

**Function** *PR-IE-CHMatching* ($\mathbb{C}, \mathbb{K}, \mathbb{M}, \mathbb{R}$)

1.  $\mathbb{T} \leftarrow \{\}$
2.  **foreach** $k$ **in** $\mathbb{K}$ **do**
3.    **foreach** $c$ **in** $\mathbb{C}$ **do**
4.      **for** $m = 1:M$ **do**
5.        **if** $\mathbb{R}[m] \neq -1$ **then**:
6.          compute the SINR of individual
            entity $c$ using Equation (4);
7.          **if** constraints (h) of Equation (13)
            is satisfied
            **then**:
8.            $x \leftarrow \min(P_{max}, \frac{\frac{P_C \cdot h_C}{\delta_{thr}} - \sigma^2}{h_{\mathbb{R}[m],c,k}^m})$;
              //calculate the upper bound of PR's
              transmitted power.
9.            $\mathbb{T} \leftarrow \mathbb{T} + \{(c, k, m, 1), x\}$;
10.           **end if**;
11.         **end if**;
12.       **end for**;
13.     **end foreach**;
14.   **end foreach**;
15.   **return** $\mathbb{T}$;

---

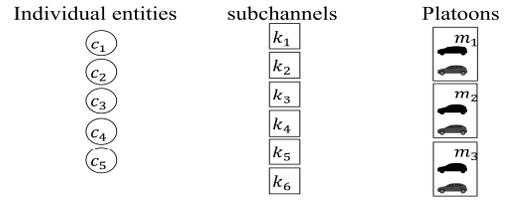

**Figure 2.** An example input of the RSPG algorithm.

| $k_1$ | $c_1$ | $c_2$ | $c_3$ | $c_4$ | $c_5$ | $k_4$ | $c_1$ | $c_2$ | $c_3$ | $c_4$ | $c_5$ |
|---|---|---|---|---|---|---|---|---|---|---|---|
| $m_1$ | 29.4 | 42.1 | 37.5 | 20.1 | 30.5 | $m_1$ | 44.8 | 30.4 | 31.4 | 41.3 | 28.4 |
| $m_2$ | 35.2 | 25.1 | 40.2 | 22.3 | 25.4 | $m_2$ | 42.6 | 33.5 | 21.1 | 36.6 | 35.6 |
| $m_3$ | 31.7 | 30.4 | 37.2 | 27.5 | 14.5 | $m_3$ | 39.6 | 27.6 | 21.6 | 24.6 | 38.6 |
| $k_2$ | $c_1$ | $c_2$ | $c_3$ | $c_4$ | $c_5$ | $k_5$ | $c_1$ | $c_2$ | $c_3$ | $c_4$ | $c_5$ |
| $m_1$ | 22.4 | 32.1 | 32.7 | 24.1 | 20.5 | $m_1$ | 35.6 | 31.4 | 27.1 | 42.6 | 22.2 |
| $m_2$ | 33.2 | 35.7 | 34.3 | 32.3 | 24.1 | $m_2$ | 44.6 | 28.6 | 39.6 | 36.6 | 27.1 |
| $m_3$ | 31.5 | 31.4 | 33.3 | 30.8 | 21.1 | $m_3$ | 40.3 | 26.6 | 31.4 | 30.6 | 27.3 |
| $k_3$ | $c_1$ | $c_2$ | $c_3$ | $c_4$ | $c_5$ | $k_6$ | $c_1$ | $c_2$ | $c_3$ | $c_4$ | $c_5$ |
| $m_1$ | 37.4 | 44.4 | 31.4 | 42.5 | 25.2 | $m_1$ | 36.1 | 38.9 | 37.5 | 30.6 | 24.6 |
| $m_2$ | 39.5 | 24.3 | 22.5 | 43.5 | 24.6 | $m_2$ | 36.6 | 22.6 | 30.6 | 33.6 | 25.7 |
| $m_3$ | 34.6 | 39.6 | 44.2 | 42.6 | 35.6 | $m_3$ | 28.4 | 42.7 | 39.9 | 35.6 | 44.6 |

(a)

| $k_1$ | $c_1$ | $c_2$ | $c_3$ | $c_4$ | $c_5$ | $k_4$ | $c_1$ | $c_2$ | $c_3$ | $c_4$ | $c_5$ |
|---|---|---|---|---|---|---|---|---|---|---|---|
| $m_1$ | | 42.1 | 37.5 | | 30.5 | $m_1$ | 44.8 | 30.4 | 31.4 | 41.3 | |
| $m_2$ | 35.2 | | 40.2 | | | $m_2$ | 42.6 | 33.5 | | 36.6 | 35.6 |
| $m_3$ | 31.7 | 30.4 | 37.2 | | | $m_3$ | 39.6 | | | | 38.6 |
| $k_2$ | $c_1$ | $c_2$ | $c_3$ | $c_4$ | $c_5$ | $k_5$ | $c_1$ | $c_2$ | $c_3$ | $c_4$ | $c_5$ |
| $m_1$ | | 32.1 | 32.7 | | | $m_1$ | 35.6 | 31.4 | | 42.6 | |
| $m_2$ | 33.2 | 35.7 | 34.3 | 32.3 | | $m_2$ | 44.6 | | 39.6 | 36.6 | |
| $m_3$ | 31.5 | 31.4 | 33.3 | 30.8 | | $m_3$ | 40.3 | | 31.4 | 30.6 | |
| $k_3$ | $c_1$ | $c_2$ | $c_3$ | $c_4$ | $c_5$ | $k_6$ | $c_1$ | $c_2$ | $c_3$ | $c_4$ | $c_5$ |
| $m_1$ | 37.4 | 44.4 | 31.4 | 42.5 | | $m_1$ | 36.1 | 38.9 | 37.5 | 30.6 | |
| $m_2$ | 39.5 | | | 43.5 | | $m_2$ | 36.6 | | 30.6 | | |
| $m_3$ | 34.6 | 39.6 | 44.2 | 42.6 | 35.6 | $m_3$ | | 42.7 | 39.9 | 35.6 | 44.6 |

(b)

**Figure 3.** (a) The SINR of each individual entity in different matchings; (b) the SINR of those matchings that can satisfy individual entity's SINR requirement.

higher than the maximum power $P_{max}$. Line 23 of the RSPG algorithm sorts the elements in set $\mathbb{T}$ based on PRV's transmitted power, i.e., the value of $x$, from high to low, i.e., descendently. Line 24 of the RSPG algorithm calls function *ResultedMatching* to find all resulted matchings for PRV's groupcasting. Function *ResultedMatching* iteratively picks up the candidate matching that has the $i_{\bar{\mathbb{T}}}$-th highest transmitted power: If entity $c$, PRV of platoon $m$ and subchannel $k$ have not been matched, then add the matching to set $\mathbb{S}$; then remove $c$, $k$ and $m$ from $\mathbb{C}$, $\mathbb{K}$ and $\mathbb{M}$ respectively in Lines 7 to 9 of function *ResultedMatching* because they have been allocated. Then, the resource allocation results can be decided according to the resulted matching set $\mathbb{S}$ of the RSPG algorithm, which shows the resource sharing among IEs, subchannels, PLVs and PRVs of all platoons.

A simple example of executing the RSPG algorithm is presented in Figures 2~8. Referring to Figure 2, let there be 5 IEs, 6 subchannels and 3 platoons. The RSPG algorithm can compute the SINR of IE $c$ using Equation (4) for every matching $(c, k, m, 0)$. Figure 3-(a) is the SINR of each IE $c_x, x = 1..5$, in different matchings, i.e., each IE $c_x, x = 1..5$, shares subchannel $k_y, y = 1..6$, with the PLV of platoon $m_z, z = 1..3$, that are calculated using Equation (4). Figure 3-(b) is the results after executing function *PL-IE-CHMatching*, i.e., excluding those matchings that can not satisfy IE's SINR requirement, which is depicted in constraint (h) of Equation (13). That is, Figure 3-(b) is the SINR of each IE for candidate matchings with the corresponding PLV of platoon $m_y, y = 1..3$, using subchannel $k_z, z = 1..6$.

Figure 4-(a) is the resulted set $\mathbb{T}$, in which each element contains a candidate matching and the correspond PLV's transmitted power. Figure 4-(b) is the resulted set $\bar{\mathbb{T}}$, which is the sorted result of $\mathbb{T}$'s elements based on PLV's transmitted power from high to low, i.e., descendently, after executing Line 3 of the RSPG algorithm. Fig 4-(c) is the resulted set $\mathbb{S}$, whose elements are selected from set $\bar{\mathbb{T}}$ and represents the resulted matching after executing function *ResultedMatching*, e.g.,

platoon $m_1$'s PLV shares subchannel $k_3$ with IE $c_2$. After executing Lines 6 to 20 of the RSPG algorithm, $\mathbb{R}$ is equal to [4, -1, 3], which means that (i) the PRV of platoon 1 and 3 is

$\mathbb{T}$

$\{(c_1, k_1, m_2, 0), 43.4\}$
$\{(c_1, k_1, m_3, 0), 40.5\}$
$\{(c_2, k_1, m_1, 0), 37.2\}$
$\{(c_2, k_1, m_3, 0), 38.4\}$
$\vdots$
$\{(c_5, k_6, m_3, 0), 41.1\}$

(a)

$\bar{\mathbb{T}}$

$\{(c_2, k_3, m_1, 0), 50.7\}$
$\{(c_3, k_3, m_3, 0), 50.4\}$
$\{(c_2, k_6, m_3, 0), 49.8\}$
$\{(c_1, k_4, m_1, 0), 49.6\}$
$\{(c_1, k_5, m_3, 0), 48.3\}$
$\{(c_1, k_3, m_2, 0), 46.1\}$
$\vdots$

(b)

$\mathbb{S}$

$\{(c_2, k_3, m_1, 0), 50.7\}$
$\{(c_1, k_5, m_3, 0), 48.3\}$
$\{(c_3, k_1, m_2, 0), 39.1\}$

(c)

**Figure 4.** (a) The resulted set $\mathbb{T}$, in which each element contains a candidate matching and the correspond PL vehicle's transmitted power; (b) the resulted set $\bar{\mathbb{T}}$, which is the sorted result of $\mathbb{T}$; (c) the resulted set $\mathbb{S}$ for PL groupcasting.

PMV 4 and PMV 3 respectively, and (ii) the transmitted power of platoon 2's PLV can reach its tail PMV, i.e., PM 5, and thus no PRV is needed for platoon 2, i.e., $R_2 = -1$.

Figure 5 depicts the configuration of PLVs and PRVs of platoons 1, 2 and 3. After that, Lines 22 to 24 of the RSPG algorithm match PRVs, IEs and subchannels using the same steps as that for matching PLV, IEs and subchannels. Since (i)

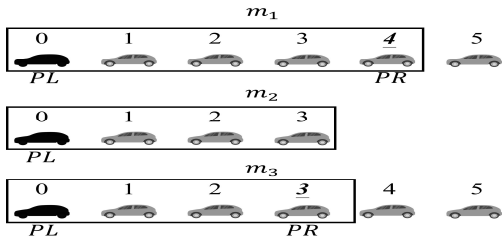

**Figure 5.** The configuration of PL and PR vehicles.

| $k_2$ | $c_4$ | $c_5$ |
|---|---|---|
| $m_1$ | 37.5 | 35.1 |
| $m_3$ | 28.8 | 33.8 |
| $k_4$ | $c_4$ | $c_5$ |
| $m_1$ | 31.5 | 29.6 |
| $m_3$ | 39.3 | 29.0 |
| $k_6$ | $c_4$ | $c_5$ |
| $m_1$ | 39.2 | 29.1 |
| $m_3$ | 33.3 | 37.2 |

(a)

| $k_2$ | $c_4$ | $c_5$ |
|---|---|---|
| $m_1$ | 37.5 | |
| $m_3$ | | 33.8 |
| $k_4$ | $c_4$ | $c_5$ |
| $m_1$ | 31.5 | |
| $m_3$ | 39.3 | |
| $k_6$ | $c_4$ | $c_5$ |
| $m_1$ | 39.2 | |
| $m_3$ | 33.3 | 37.2 |

(b)

**Figure 6.** (a) The SINR of each IE in different matchings; (b) the SINR of each IE for candidate matchings with the corresponding PR.

IEs 1, 2 and 3 can share subchannels with PLVs of platoons 1, 2 and 3, respectively and (ii) subchannels 1, 3 and 5 have been used, only IEs 4 and 5 can potentially share subchannels 2, 4 and 6 with PRVs. Figure 6-(a) is the SINR of each IE $c_x$, $x = 4, 5$, in different matchings, i.e., each IE $c_x$, $x = 4, 5$, shares subchannel $k_y$, $y = 2, 4, 6$, with PRV of platoon $m_z$, $z = 1, 3$, that are calculated using Equation (4). Figure 6-(b) is the results after executing function *PR-IE-CHMatching*, i.e., excluding those matchings that cannot satisfy IE's SINR requirement, which is depicted in constraint (13f) of Equation (13). That is, Figure 6-(b) is the SINR of each IE for candidate matchings with the corresponding PRV of platoon $m_y$, $y = 1, 3$, using subchannel $k_z$, $z = 2, 4, 6$. Figure 7-(a) is the resulted set $\mathbb{T}$, in which each element contains a candidate matching and the correspond PRV's transmitted power. Figure 7-(b) is the resulted set $\overline{\mathbb{T}}$, which is the sorted result of $\mathbb{T}$'s elements based on PRV's transmitted power from high to low, i.e., descendently, after executing Line 23 of the RSPG algorithm. Fig 7-(c) is the resulted set $\mathbb{S}$, whose elements are selected from set $\overline{\mathbb{T}}$ and represents the resulted matching after executing Line 24 of the RSPG algorithm, e.g., platoon $m_3$'s PRV shares subchannel $k_4$ with IE $c_4$.

Figure 8 depicts the resulted matching set $\mathbb{S}$ after executing the RSPG algorithm, which is represented as a tripartite matching graph.

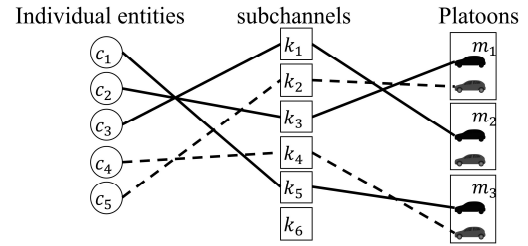

|  $\mathbb{T}$  |  $\overline{\mathbb{T}}$  |  $\mathbb{S}$  |
|---|---|---|
| $\{(c_4, k_2, m_1, 1), 33.4\}$ | $\{(c_4, k_4, m_3, 1), 37.6\}$ | $\{(c_2, k_3, m_1, 0), 50.7\}$ |
| $\{(c_4, k_2, m_3, 1), 29.7\}$ | $\{(c_4, k_2, m_1, 1), 37.5\}$ | $\{(c_1, k_5, m_3, 0), 48.3\}$ |
| $\{(c_5, k_2, m_1, 1), 32.8\}$ | $\{(c_5, k_2, m_3, 1), 35.4\}$ | $\{(c_3, k_1, m_2, 0), 39.1\}$ |
| $\{(c_5, k_2, m_3, 1), 34.1\}$ | $\{(c_4, k_4, m_1, 1), 34.1\}$ | $\{(c_4, k_4, m_3, 1), 37.6\}$ |
| $\vdots$ | $\vdots$ | $\{(c_5, k_2, m_1, 1), 32.8\}$ |
| $\{(c_5, k_6, m_3, 1), 35.6\}$ | | |

(a)                (b)                (c)

**Figure 7.** (a) The resulted set $\mathbb{T}$, in which each element contains a candidate matching and the correspond PR vehicle's transmitted power;
(b) the resulted set $\overline{\mathbb{T}}$, which is the sorted result of $\mathbb{T}$; (c) the resulted set $\mathbb{S}$ for PL and PR groupcasting.

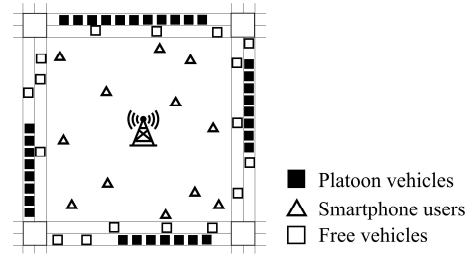

**——** : matching with the PL vehicle  **-- --** : matching with the PR vehicle

**Figure 8.** The example's resulted tripartite matching graph.

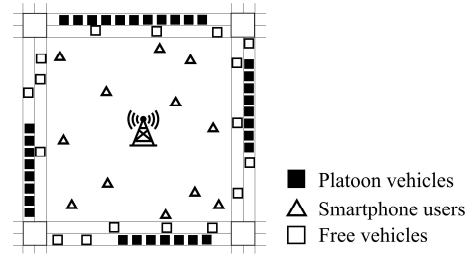

**Figure 9.** An illustrated configuration of the simulation.

■ Platoon vehicles
△ Smartphone users
□ Free vehicles

perpendicular manner. The simulation parameters and their values are depicted in Table 1.

The compared methods with the proposed method, i.e., the proposed RSPG algorithm, for relay selection are as follows: (1) The proposed method. (2) The centralized method proposed in [4], for which the BS always selects a PRV that can minimize both the transmitted power of the PLV and the transmitted power of the PRV in a platoon. (3) The No Relay method, in which PLV's groupcasting has the responsibility to transmit PLV's messages to platoon's tail vehicle using the corresponding transmitted power. Additionally, the allocated subchannel for PLV's groupcasting can be shared with one IE, under the condition of the SINR of PLV's groupcasting being not lower than the SINR's minimum threshold. The adopted performance metrics that are used for comparison are as follows:

(I) Platoon's transmission latency (ms): It denotes the average end-to-end transmission time for different cases. For groupcasting, it includes two pieces of transmissions: (1) PLV groupcasts messages to (a) the PRV and (b) the PMVs that are

## VI. Performance Evaluation

In this Section, the performance evaluation of the proposed method is presented. The simulation environment and the performance metrics are explained at first. Then, the simulation results are presented to validate the proposed method.

### 6.1. The Simulation Environment

To evaluate the performance of the proposed method, an urban environment has been modeled, which is depicted in Figure 9. The urban environment simulates an urban street block covered by a single cell, where the BS is located at the center of the block. Four roads surround the BS in a between PLV and PRVs and (2) PRV groupcasts messages to the PMVs that are between PRV and the last PMV.

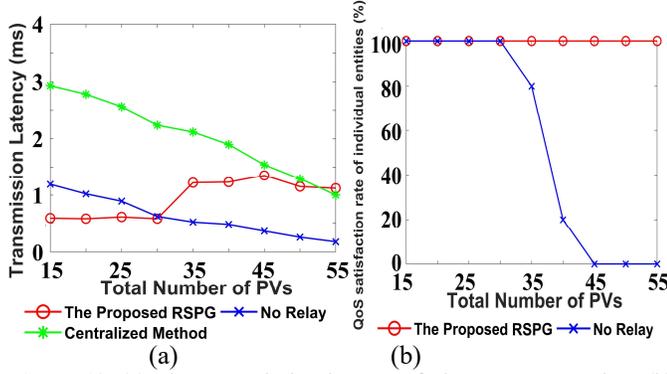

**Figure 10.** (a) The transmission latency of platoon groupcasting; (b) the QoS's satisfaction rate of IEs.

(II) QoS's satisfaction rate of IEs: The QoS's satisfaction rate of IEs denotes the number of IEs whose SINRs are higher than the minimum threshold divided by the number of IEs that have shared subchannel with platoon vehicles.

(III) Number of allocated subchannels: The number of subchannels that are allocated to PLVs'/PRVs' groupcasting.

(IV) Spectral efficiency (SE) (bps/Hz): The bit rate that can be used over a transmission Hz in the system. It is calculated as dividing the transmission bit rate by the allocated subchannels' amount of transmission bandwidth, which is in the unit of Hz. The spectral efficiency of PLV's and PRV's groupcasting in a platoon are evaluated in this work.

### 6.2. The Simulation Results

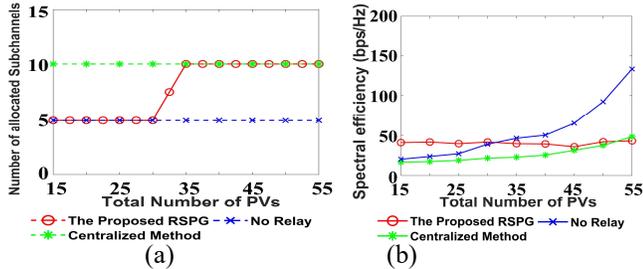

**Figure 11.** (a) The number of allocated subchannels for platoon's groupcasting; (b) the spectral efficiency of platoon's groupcasting.

Let the number of IEs be 65, and the number of platoons be 5 (M = 5); the number of IEs be bigger than the total number of platoon vehicles of these 5 platoons because the simulation is assumed to be in the urban scenario; the number of subchannels that can be assigned be 65 subchannels (K = 65).

Figure 10-(a) depicts the transmission latency of PLV's and PRV's groupcasting. When the number of platoon vehicles is equal to or smaller than 30, in which condition all platoons in the proposed RSPG and the No Relay methods don't need PRV, (i) the RSPG method remains low transmission latency and (ii) the No Relay method's latency decreases when the number of platoons' vehicles is increased. The reason is as follows. Using the proposed RSPG, PLV uses the transmitted power that can satisfy both the SINR requirements of PLV and the corresponding IE, which shares its allocated subchannel with PLV, and thus it can keep the stable transmission latency. Using the No Relay method, PLV increases its transmitted power to transmit messages to platoon's tail vehicle when the number of platoon's vehicles is increased, which results in the higher SINR. The higher SINR thus can increase the transmission rate, which, in turn, results in the lower transmission latency. When the number of platoon vehicles is more than 30, the platoon using the proposed RSPG needs to pick a PRV to forward PLV's messages. As a result, each message needs to be received completely in the PRV at first; then the received message is re-groupcasted by the PRV. Therefore, the proposed RSPG's transmission latency becomes more than two times higher than the transmission latency on the conditions of the number of platoon vehicles being equal to or smaller than 30. The Centralized Method always has the higher transmission latency than the other two methods when the total number of platoon

Table 1: Parameters and their values used in the simulation.

| Parameters | Values |
|---|---|
| Radius of the BS | 1km |
| BS's antenna height | 25m |
| BS's antenna gain | 8 dBi |
| The distance from the BS to road | 100 m |
| Vehicle's antenna height | 1.5 m |
| Road lanes | 2 |
| Lane width of the road | 4 m |
| SINR threshold for the receiver ($SINR_{thr}$) | 5 dB |
| QoS requirement of individual entities | 0.5 bps/Hz |
| Number of platoons ($M$) | 5 |
| The maximum platoon size | 11 |
| Number of individual entities | 65 |
| Vehicle's maximum transmitted power for groupcasting | 30 dBm |
| Ies' maximum transmitted power | 30 dBm |
| Bandwidth | 10 MHz |
| Carrier frequency | 2 GHz |
| Fading factor ($\alpha$) | 3 |
| Noise power spectrum density ($\sigma^2$) | -114 dBm |
| Size of a data packet ($\lambda$) | 300 Bytes |
| Pathloss model for individual cellular network users | 128.1 + 37.6 $\log_{10}(d)$ [10] |
| Pathloss model for platoon vehicles' communications | *LOS WINNER + B1* [10] |

vehicles is smaller than 55. The reason is that it needs to pick a PRV no matter how long the platoon is and thus it needs to experience the transmission latency of the PLV and the transmission latency of the PRV. Additionally, the Centralized method starts from high latency and then the latency becomes smaller when the number of platoon vehicles is increased. The reason is that the transmitted power of both PLVs' and PRVs' groupcasting is increased when the total number of vehicles in platoons is increased. The increased power results in the higher SNR, which, in turn, results in the higher transmission rate and lower transmission latency.

Referring to Figure 10-(b), the QoS's satisfaction rate of IEs using the No Relay method starts decreasing rapidly when the total number of vehicles in platoons is bigger than 30. The reason is that increasing PLVs' transmitted power makes IEs' interference be increased using the No Relay method, which leads to IEs' SINR to be lower than the minimum requirement.

Figure 11-(a) depicts the number of allocated subchannels for PLV's and PRV's groupcasting. Referring to Figure 11-(a), the proposed RSPG does not need to pick a PRV when the number of platoon vehicles is equal to or smaller than 30. When the number of platoon vehicles is bigger than 30, the proposed RSPG method needs to allocate one more subchannel for PRV's groupcasting in each platoon. Thus, the number of allocated subchannels in the situations of platoon vehicles' number being greater than 30 is twice of the number of allocated subchannels in the situations of platoon vehicles' number being equal to or smaller than 30. The number of subchannels used in the No Relay method is the same as the number of platoons because it uses one subchannel for PLV's groupcasting for each platoon. On the other hand, since the Centralized methods always needs one subchannel for PLV's groupcasting and one subchannel for PRV's groupcasting, the number of allocated subchannels of using the Centralized method for one platoon is 2, which is twice of that of using the No Relay method.

Figure 11-(b) depicts the spectral efficiency of those subchannels allocated to PLVs/PRVs. The proposed RSPG method remains to the similar spectral efficiency when the total number of platoon vehicles is increased from 15 to 55. The reason that the proposed RSPG method achieves stable spectral efficiency is as follows. Although the proposed RSPG method uses more subchannels when the total number of platoon vehicles is bigger than 30, the proposed RSPG considers all IEs that can share subchannels with PLVs and PRVs to find the suitable transmitted power, which makes PLV and PRV adjust the transmitted power to have the similar SINR, and thus, results in the similar transmission rate even if the number of platoon vehicles is increased. Thus, the spectral efficiency remains similar. The No Relay method's spectral efficiency increases when number of platoon vehicles increases because the transmitted power of PLVs increases, which, in turn, increases the SINR and thus increases the transmission rate. As a result, the spectral efficiency increases. The spectral efficiency of the Centralized method slightly increases when the number of platoon vehicles increases because the transmitted power of PLVs and PRVs increases, which, in turn, increases the SINR and thus increases the transmission rate. As a result, the spectral efficiency increases.

## VII. Conclusion

In this work, the resource allocation and resource sharing for multi-platoon communications based on groupcasting's and unicasting's reliability of platoons has been studied. The proposed RSPG method has the resource allocation and resource sharing for PLV's/PRVs' groupcasting. For PLV's/PRV's groupcasting, a tripartite matching problem that matches (i) an allocated subchannel, (ii) a PLV or a PRV and (iii) an IE is formulated and solved using the proposed RSPG method. The RSPG method maximizes PLV's/PRV's transmitted power that can make the SINR of the IE, which shares the subchannel with the PLV/PRV, be higher than the minimum threshold. The RSPG method's tripartite matching result denotes the resource allocation for PLVs/PRVs. The simulation results have shown that the proposed methods have the better performance on (1) QoS's satisfaction rate, (2) the number of allocated subchannels for platoon vehicles and (3) the spectral efficiency than the other compared methods in the urban scenario. The future work can consider to apply the resource allocation of multi-platoon communications from single-cell to multi-cell, where intra-cell and inter-cell interferences need to be considered.

**Acknowledgments** This work was supported in part by the Ministry of Science and Technology (MOST), Taiwan, under the grant number MOST 111-2221-E-006-077-MY3; and in part by the Funds for Science and Technology Development of the University of Danang under Project B2021-DN02-06.